\documentclass[aps,prx,reprint,noeprint,floatfix]{revtex4-2}
\usepackage[utf8]{inputenc}
\usepackage{xcolor}
\usepackage{amsmath}
\usepackage{amssymb}
\usepackage{physics}
\usepackage{graphicx}
\usepackage{parskip}
\usepackage{hyperref}
\usepackage{varioref}       
\usepackage{cleveref}     

\crefname{table}{table}{tables}
\Crefname{table}{Table}{Tables}
\crefname{figure}{fig.}{figs.}
\Crefname{figure}{Fig.}{Figs.}
\crefname{section}{sect.}{sects.}
\Crefname{Section}{Sect.}{Sects.}

\definecolor{nicecolor}{rgb}{0.03, 0.27, 0.49}
\hypersetup{colorlinks=true,allcolors = nicecolor,linktocpage=true}

\usepackage{mathptmx}

\newcommand{\commie}[1]{}
\newcommand{\ads}{\mathrm{AdS}}

\newenvironment{eqaed}
    {\begin{equation}
    \begin{aligned}
    }
    { 
    \end{aligned}
    \end{equation}
    \ignorespacesafterend
    }

\begin{document}

\begin{titlepage}

\begin{flushright}
MPP-2025-129
\end{flushright}

\title{\vspace{-32pt}\bf Dark bubble cosmology and the equivalence principle}
\author{$^1$Ivano Basile}
\email{ibasile@mpp.mpg.de}
\author{$^2$Alessandro Borys}
\email{alessandro.borys@dfa.unict.it}
\author{$^1$Joaquin Masias}
\email{jmasias@mpp.mpg.de}
\affiliation{$^1$Max-Planck-Institut f\"ur Physik (Werner-Heisenberg-Institut)\\ Boltzmannstraße 8, 85748 Garching, Germany}
\affiliation{$^2$Department of Physics, University of Catania, and INFN\\Via Santa Sofia 64, I-95123 Catania, Italy}

\begin{abstract}

    \noindent The main goal of string phenomenology is to find realistic models of particle physics and cosmology within string theory. Dark bubble cosmology is an alternative to string compactifications, where our universe lives on a bubble expanding in a higher-dimensional spacetime. This construction is inherently non-supersymmetric and can yield a four-dimensional realistic cosmology, where radiation behaves as expected due to its coupling to higher-dimensional fields. We study the coupling of the electroweak and strong sectors to the induced braneworld gravity via the same mechanism. While the electroweak sector is unaffected, the gravitational and inertial masses of the proton differ significantly, severely violating measurements of the equivalence principle.
    
    \end{abstract}

\maketitle

\end{titlepage}
\setcounter{page}{1}

\section{Introduction}
The dark bubble model \cite{Banerjee:2018qey, Banerjee_2019, Banerjee:2020wix,Banerjee:2021yrb,Danielsson:2021tyb,Banerjee:2022ree,Danielsson:2022lsl,Danielsson:2022fhd, Basile:2023tvh, Banerjee:2023uto,danielsson2024chargednariaiblackholes, Basile:2025lek} offers an alternative to standard compactification attempts to obtain (quasi-)de Sitter vacua in string theory \footnote{See also \cite{Muntz:2025ltu} for a related proposal with end-of-the-world branes.}. According to this proposal, our universe lives on a 3-brane on the wall of a bubble which interpolates between two bulk 5d non-supersymmetric anti-de Sitter (AdS) vacua with slightly different energy density. As the bubble expands, 4d observers confined to the bubble effectively perceive an expanding Friedmann-Lema\^{i}tre-Robertson-Walker (FLRW) universe.

Thus far, two string embeddings of the dark bubble scenario have been proposed \cite{Danielsson:2022lsl, Danielsson:2023alz, Basile:2025lek}. The former employs rotating branes in the AdS$_5 \times S^5$ geometry of type IIB string theory, breaking supersymmetry and generating a positive vacuum energy. The latter takes place in type 0'B string theory \cite{Sagnotti:1995ga, Sagnotti:1996qj} and realizes a dynamical dark energy which contradicts observations \cite{Basile:2025lek}. 

Although the dark bubble scenario is not a standard 4d effective field theory \footnote{As such, it is expected to violate several swampland conditions \cite{Anastasi:2025puv}, although the AdS bulk does not.}, gravitational waves \cite{Danielsson:2022fhd} and abelian gauge fields \cite{Basile:2023tvh, Basile:2025lek} behave correctly due to their coupling to bulk fields. In this letter we study the same mechanism non-abelian gauge fields, finding a different low-energy coupling to braneworld gravity which violates the equivalence between inertial and gravitational mass. We show that for the electroweak sector of the Standard Model there is no violation at tree level, while the strong sector shows a \textit{significant deviation for the proton}. Comparing our results with data, we conclude that the dark bubble scenario, at least as currently formulated, is incompatible with the Standard Model.
    
In Section \ref{review} we briefly present the main features of the model and its induced gravity. In Section \ref{section2} we introduce gauge fields, extending previous work to the non-abelian case. In contrast to the abelian case, non-abelian fields do not couple to braneworld gravity as in 4d effective field theory.
In Section \ref{section3} we show that the tree-level masses of W and Z bosons are unaffected, while in Section \ref{section4} we show that gluons lead to large violations of the equivalence principle for the proton mass.

\section{The dark bubble scenario}\label{review}
In contrast with constructions \`{a} la Randall-Sundrum, the dark bubble scenario is an inside-outside construction where two $\ads_5$ vacua have slightly different cosmological constants. At distances longer than the AdS radii, the 4d Einstein equations follow from the junction condition across the brane via the Gauss-Codazzi relation \cite{Banerjee:2018qey, Banerjee:2019fzz}. The induced geometry is sourced by the energy-momentum tensor of the brane itself (which for empty branes acts as a cosmological constant) and by contributions from the AdS bulk. Upon adding matter to the brane, the full backreacted bulk solution leads to a net \textit{positive} energy density, taking into account the extrinsic curvature through the junction condition. Stretched strings pulling on the brane look like particles in 4d, whose mass is related to the string tension by the bulk geometry in such a way that only 5d AdS produces a constant mass along the expansion of the brane \cite{Basile:2020mpt}. Similarly, radiation is implemented by 5d black holes.

As outlined above, the energy-momentum tensor of the brane sources a jump in the extrinsic curvature. In terms of the proper time $\tau$ on the brane located at $r=a(\tau)$ in global AdS coordinates, the induced metric on the brane has the FLRW form
\begin{equation}
    ds^2_\text{brane}=-d\tau^2+a(\tau)^2 d\Omega_3^2 \, .
\end{equation}
The Israel junction conditions \cite{Israel:1966rt} then imply that the brane tension $\sigma$ be
\begin{equation}
        \sigma=\frac{3}{8\pi G_5}\left(\sqrt{k_-^2+\frac{1+\dot a^2}{a^2}}-\sqrt{k_+^2+\frac{1+\dot a^2}{a^2}}\right),
    \end{equation}
where the cosmological constants inside and outside the shell are given by $\Lambda_\pm=-6k_\pm^2=-6/L_\pm^2$ and satisfy $\Lambda_-<\Lambda_+<0$ (or $k_->k_+$). The vacuum with higher energy can then decay through the nucleation of a spherical bubble, which depending on the specific construction may be a Coleman-de Luccia or Brown-Teitelboim bubble; the 3-brane is the wall of this bubble.

When $k_- - k_+ \ll k_\pm$ the tension is slightly below the critical value $\sigma_\text{cr} \equiv \frac{3}{8\pi G_5}(k_--k_+)$, and the scale factor satisfies
\begin{equation}
    \frac{\dot a^2}{a^2}\sim -\frac{1}{a^2}+\frac{8\pi G_4}{3}\Lambda_4 \, ,
\end{equation}
where 
\begin{equation}
        \Lambda_4=\frac{3(k_--k_+)}{8\pi G_5}-\sigma \, , \qquad G_4 = \frac{2k_-k_+}{k_--k_+} \, G_5 
\end{equation}
are the (positive) cosmological constant and Newton constant in 4d. Since the 4d cosmological constant is proportional to $\sigma_\text{cr} - \sigma$, we see how a slightly subcritical brane is essential to realize scale separation of the braneworld physics with respect to the bulk.

All in all, the induced 4d cosmology on the brane is not a standard 4d effective field theory. In other words, the physics is inherently higher-dimensional, and the separation of scales follows from the interplay of various higher-dimensional ingredients, rather than a standard compactification.

\section{Non-abelian fields in the dark bubble}\label{section2}
We now discuss how gauge fields couple to braneworld gravity. Stringy realizations of the dark bubble contain 2-form fields $B$ and $C$, which couple to the worldvolume of the D3-branes on the wall of the dark bubble. This coupling induces a backreaction in the bulk, which in turn modifies the induced geometry on the brane.

In the abelian case \cite{Basile:2023tvh, Basile:2025lek} the $B$-field couples to the Maxwell field $F$ on the D3-brane worldvolume according to the (Dirac-)Born-Infeld (DBI) action
\begin{equation}
    S_\text{D3}=-T_3 \int d^4x\sqrt{-\det(g_4+\tau \mathcal{F})} \, ,
\end{equation}
where $\tau=2\pi\alpha'$ with $\alpha'\equiv l_s^2$ the (reciprocal of the) string tension. The D3-brane tension is $T_3=\frac{1}{(2\pi)^3\alpha'^2g_s}$ in terms of the string coupling $g_s$, and the combination $\tau\mathcal{F}=\tau F+B$ contains the Maxwell curvature $F$ together with the $B$-field. At low curvatures with respect to the string scale
\begin{eqaed}
    S_\text{D3} &\sim -\int d^4x\sqrt{-\det(g_4)}\left(T_3+\frac{1}{4g^2}\mathcal{F}_{\mu\nu}\mathcal{F}^{\mu\nu}\right), 
    \label{eq:DBI}
\end{eqaed} 
where $g^2\equiv2\pi g_s$ is the gauge coupling. The bulk-brane dynamics then arise varying the action
\begin{equation}
     S_5=\frac{1}{2\kappa_5}\int d^5x\sqrt{-\det(g_5)}\left(R-\frac{1}{12g_s}H^2 \right) + S_\text{D3} \, ,
\end{equation}
where $H=dB$ is the curvature of $B$. It follows that the $B$-field induces a discontinuity across the brane \cite{Basile:2023tvh}
\begin{equation}
\Delta H^{r\mu\nu}\big|_{r=a}=\frac{8 G_5 ka}{\pi\alpha'}\mathcal{F}_{\mu\nu}\big|_{r=a} \, , \label{eq:Delta}
\end{equation}
which means that electromagnetic waves on the brane source the $B$-field in the bulk. The Israel junction conditions then yield
the 4d Einstein equations with the correct Maxwell stress tensor \cite{Basile:2023tvh}.

Ramond-Ramond (R-R) potentials $C_n$ also couple to the worldvolume gauge field via the Chern-Simons term. For a D$p$-brane sweeping a worldvolume $W$, it reads \cite{Morales:1998ux}
\begin{equation}
S_\text{CS}=\mu_p\sum_n \int_{W}C_n\wedge\sqrt{\frac{\hat A(TW)}{\hat A(NW)}}\wedge e^{\tau \mathcal{F}} \, ,
\end{equation}
where the minimal charge $\mu_p=\frac{1}{(2\pi)^p l_s^{p+1}}$. For a D3-brane, the relevant couplings are
\begin{equation}
    S_\text{CS}=\mu_3\int_{W}(C_4+\tau \mathcal{F}\wedge C_2) \, .
\end{equation}
Including the R-R fields in the bulk action, the field equations are \cite{Basile:2025lek}
\begin{eqaed}
\partial_rH^{r\mu\nu}
&= \frac{2G_5 kr}{\pi^2\alpha'^2}\left[\tilde C^{\mu\nu}+2\tau\mathcal{F}^{\mu\nu}\right]\delta(r-a(\eta)) \, \, \\ 
\partial_rF^{r\rho\sigma} 
&= \frac{2G_5 kr}{\pi\alpha'}\mathcal{\widetilde F}^{\rho\sigma}\delta(r-a(\eta)) \, ,
\end{eqaed}
where $\mathcal{\widetilde F}$ is the dual electromagnetic field-strength. Since $\tilde{C}$ vanishes on the brane, we obtain the discontinuities
\begin{equation}
\Delta H
\sim \mathcal{F} \hspace{1 cm} 
\Delta F_3
\sim \widetilde{\mathcal{F}}.
\end{equation}
We now generalise this construction to non-abelian worldvolume gauge fields $\tau \mathcal{F}=B\mathbb{I}+\tau F_a T^a$. The indices $a$ refer to a $SU(N)$ Lie algebra. We follow the symmetrized trace prescription \cite{Tseytlin:1997csa, Myers:1999ps} in the absence of scalar fields, which produces a valid action up to order $\mathcal{F}^4$ \cite{Hashimoto:1997gm, Sevrin:2001ha}. Then, \cref{eq:DBI} becomes 
\begin{eqaed}
    S_\text{D3} \sim - \int d^4x \, \text{STr} \sqrt{-\det(g_{\mu\nu}+\tau \mathcal{F_{\mu\nu}})} \, .
\end{eqaed}
Up to quartic order,
\begin{eqaed}
&\text{STr}\left[\mathcal{F}_{\mu\nu}^2-\frac{1}{2}\tau^2(\mathcal{F}^4-\frac{1}{4}(\mathcal{F}^2)^2)\right] \sim \text{Tr}[\mathcal{F}_{\mu\nu}\mathcal{F}^{\mu\nu}] \, \\ 
& -\frac{\tau^2}{3}\text{Tr}[\mathcal{F}_{\mu\nu}\mathcal{F}^{\alpha\nu}\mathcal{F}^{\mu\beta}\mathcal{F}_{\alpha\beta} + \frac{1}{2}\mathcal{F}_{\mu\nu}\mathcal{F}^{\alpha\nu}\mathcal{F}^{\alpha\beta}\mathcal{F}_{\mu\beta}\, \\ 
&-\frac{1}{4}\mathcal{F}_{\mu\nu}\mathcal{F}^{\mu\nu}\mathcal{F}_{\alpha\beta}\mathcal{F}^{\alpha\beta}-\frac{1}{8}\mathcal{F}_{\mu\nu}\mathcal{F}_{\alpha\beta}\mathcal{F}^{\mu\nu}\mathcal{F}^{\alpha\beta}] \, . 
\end{eqaed}
The resulting discontinuity in the $B$-field is proportional to a cubic term, schematically $\Delta H \propto \text{Tr}(F^3)$. This is because $\text{Tr}F=0$ in the non-abelian sector. Even without a specific prescription for a non-abelian DBI action, one can see that $\Delta H$ cannot be sourced by a term linear in $F$. 

This difference has profound implications. The discontinuity correspond to the induced energy-momentum tensor in the 4d Einstein equations, whose brane contribution has the wrong sign. In the abelian case, the  backreaction from the bulk $B$-field contributes the correct sign with a factor of two, resulting in the correct energy-momentum tensor $2T_{\mu \nu} - T_{\mu \nu} = + T_{\mu \nu}$. In the non-abelian sector, the minus sign from the brane contribution cannot be compensated by the higher-order terms, which are subleading in $\alpha'$.

Analogous considerations hold for R-R fields, including the axion $F_1 = dC_0$. The non-abelian CS term yields \cite{Basile:2025lek},
\begin{eqaed}
    S_\text{D3} \sim \text{Tr} \left(\mu_3\int_{W}C_4+ C_2 \wedge \tau \mathcal{F}+ \frac{1}{2}C_0 \wedge \tau^2\mathcal{F}\wedge\mathcal{F}\right) \, ,
\end{eqaed}
from which $\Delta F_1 \propto \tau^2 \text{Tr}(F\wedge F)$.

\section{Electroweak gauge bosons}\label{section3}
As we discussed above, at low energies, the minimal coupling of non-abelian gauge fields to the induced gravity comes with a negative sign. In the $SU(2)_L\times U(1)_Y$ electroweak sector of the Standard Model, the energy-momentum tensor thus reads
\begin{eqaed}
T^{\mu \nu}=T^{\mu \nu}_{\phi}-T^{\mu \nu}_{SU(2)}+T^{\mu \nu}_{U(1)} \, ,
\end{eqaed}
where the first term is the Higgs contribution. We can diagonalize the mass contributions to $T^{00}$ to find gravitational masses at tree level. The non-abelian covariant derivative is
\begin{eqaed}
    D_\mu=\partial_\mu-igA_\mu^a\tau^a-\frac{i}{2}g'B_\mu \, ,
    \label{eq:D}
\end{eqaed} 
where $A_\mu^a$ and $B_\mu$ are the $SU(2)$ and $U(1)$ gauge fields with corresponding gauge couplings $g,g'$. In the Higgs phase, $\phi$ acquires an expectation value of the form
\begin{eqaed}
    \langle\phi\rangle=\frac{1}{\sqrt{2}}\begin{pmatrix}
 0 \\
v
\end{pmatrix}.
\end{eqaed}
The energy density of perturbations around the vacuum in the temporal gauge $A^0=0$ then simplifies to
\begin{eqaed}
  T_{00}=-\frac{v^2}{32}\left(g^2(A_\alpha^1)^2+g^2(A_\alpha^2)^2+(-g A_\alpha^3+g'B_\alpha)^2\right) \, .
\end{eqaed}
To compute gravitational masses, we diagonalize the matrix
\begin{eqaed}
\begin{pmatrix}
\frac{v^2g^2}{4} & 0 & 0 & 0 \\
0 & \frac{v^2g^2}{4} & 0 & 0\\
0 & 0 & \frac{v^2g^2}{4} &- gg' \\
0 & 0 & -gg' & \frac{v^2g'^2}{4}\\
\end{pmatrix}.
\end{eqaed}
The resulting spectrum is the following:
\begin{itemize}
    \item $W_\alpha^{\pm}=\frac{1}{\sqrt{2}}(A_\alpha^1\mp iA_\alpha^2)$, with $m_W=g\frac{v}{2}$,
    \item $Z_\mu^0=\frac{1}{\sqrt{g^2+g'^2}}(gA_\alpha^3-g' B\alpha)$, with $m_Z=\sqrt{g^2+g'^2}\frac{v}{2}$,
    \item $A_\mu=\frac{1}{\sqrt{g^2+g'^2}}(g'A_\alpha^3+g B\alpha)$, with $m_A=0$.
\end{itemize}
The tree-level gravitational masses of the electroweak gauge bosons coincide with their inertial masses as in the Standard Model. We thus turn to the strong sector to find deviations from the equivalence principle.
    
\section{Gluons and hadrons}\label{section4} 

\begin{figure}
    \centering
    \includegraphics[width=0.9\linewidth]{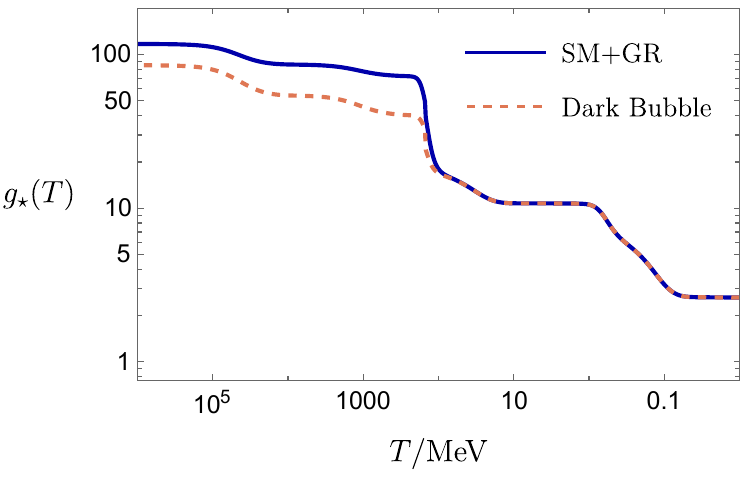}
    \caption{Number of effective relativistic degrees of freedom according to the standard thermal history of the universe (blue) and to the dark bubble scenario (orange).}
    \label{fig:thermalhist}
\end{figure}

Since the strong sector confines and is strongly coupled at low energies, we first consider the only moment in the thermal history of the universe in which one finds free gluons, the quark-gluon plasma (QGP) phase present at temperatures above the hadronization scale ($\sim 150$ MeV). If the negative contribution of gluons to the energy density takes over the total energy density $\rho$ of the plasma, this could radically modify the known thermal history. Indeed, during thermal equilibrium the Hubble parameter is $H^2 \simeq \rho$. If the total energy density became negative, the thermal history would radically differ, even breaking the underlying assumption of FLRW geometry.  During thermal equilibrium, the relative contribution of gluons can be determined by the number of degrees of freedom $g_{g}=8\times 2=16$ which comes from the $SU(3)$ degrees of freedom and helicities. Just before the QCD phase transition we also have quarks ($u, d, s$), photons, charged leptons ($e, \mu$) and neutrinos, which contribute
\begin{eqaed}
  g_{q}=36,\,\, \,   g_{\gamma}=2,\,\, \,g_{l}=8,\,\,\, g_{\nu}=6 \, .
\end{eqaed}
Since fermions contribute to the total number of relativistic degrees of freedom with a $7/8$ factor compared to bosons,
\begin{eqaed}
    g_{\star}=\sum_i g_{i,b}+\dfrac{7}{8}\sum_i g_{i,f} \, ,
\end{eqaed}
the total number of relativistic degrees of freedom before the QCD phase transition is $g_{\star}=61.5$. Then, given a total energy budget $\rho$, each particle in thermal equilibrium contributes as
\begin{eqaed}
    \rho_i=\dfrac{g_i}{g_{\star}} \, \rho \, ,
\end{eqaed}
and in particular gluons contribute a fraction of the total energy
\begin{eqaed}
    \rho_g =\dfrac{g_g}{g_{\star}}\rho\simeq 0.26 \rho \, .
\end{eqaed}
During the thermal evolution, under the assumption of isotropy, we have a FLRW geometry with Hubble scale
\begin{eqaed}
    H^2\simeq \dfrac{\rho}{M_{Pl}^2}\simeq\dfrac{T^4}{M_{Pl}^2} \, .
\end{eqaed}
If gluons coupled to gravity with the opposite sign, they would effectively contribute as negative degrees of freedom, replacing $g_{\star}$ with
\begin{eqaed}
    \tilde{g}_{\star}= g_{\star}-2 g_{g}=29.5 \, .
\end{eqaed}
This is depicted schematically in \cref{fig:thermalhist}.
Had gluons  dominated the degrees of freedom during the early universe, this would have inevitably led to a radically different thermal history (in particular giving a negative total energy density, breaking the assumption of FLRW). It is however much harder to experimentally probe a decreased number of degrees of freedom \textit{only} before QCD phase transition. Since very strongly constrained observables such as those from Big Bang Nucleosynthesis and  Neutrino Decoupling are sensitive to the number of relativistic degrees of freedom during those epochs, long after confinement takes place. The fact that the universe was opaque before photons decoupled to produce the Cosmic Microwave Background (CMB) also means that we cannot directly observe this epoch, and we would require further assumptions (e.g. additional particle content) to engineer a first-order phase transition that could lead to an imprint on the CMB. Therefore, we now turn to the low-energy regime of QCD. We will use lattice computations and data to show that gluons coupling with a negative sign to gravity necessarily lead to large violations of the equivalence principle via the gravitational mass of the proton.

\subsection{Tests of the Weak Equivalence Principle}

Einstein's Equivalence Principle (EEP) is a basic tenet of General Relativity, asserting that the laws of physics are the same in all locally inertial frames. A key component of the EEP is the Weak Equivalence Principle (WEP), which states that all test bodies fall at the same rate in a gravitational field, independently of their mass or composition. This implies the universality of free fall and leads to the idea that gravity is a manifestation of spacetime curvature. While the EEP can be violated in alternative theories (e.g. due to non locality or non metricity) \cite{Damour:1994zq,Olmo:2006zu,Will:2014kxa}, one expects the WEP to always hold. In particular, inertial and gravitational masses should agree.
         
\begin{figure}
    \centering
    \includegraphics[width=0.9\linewidth]{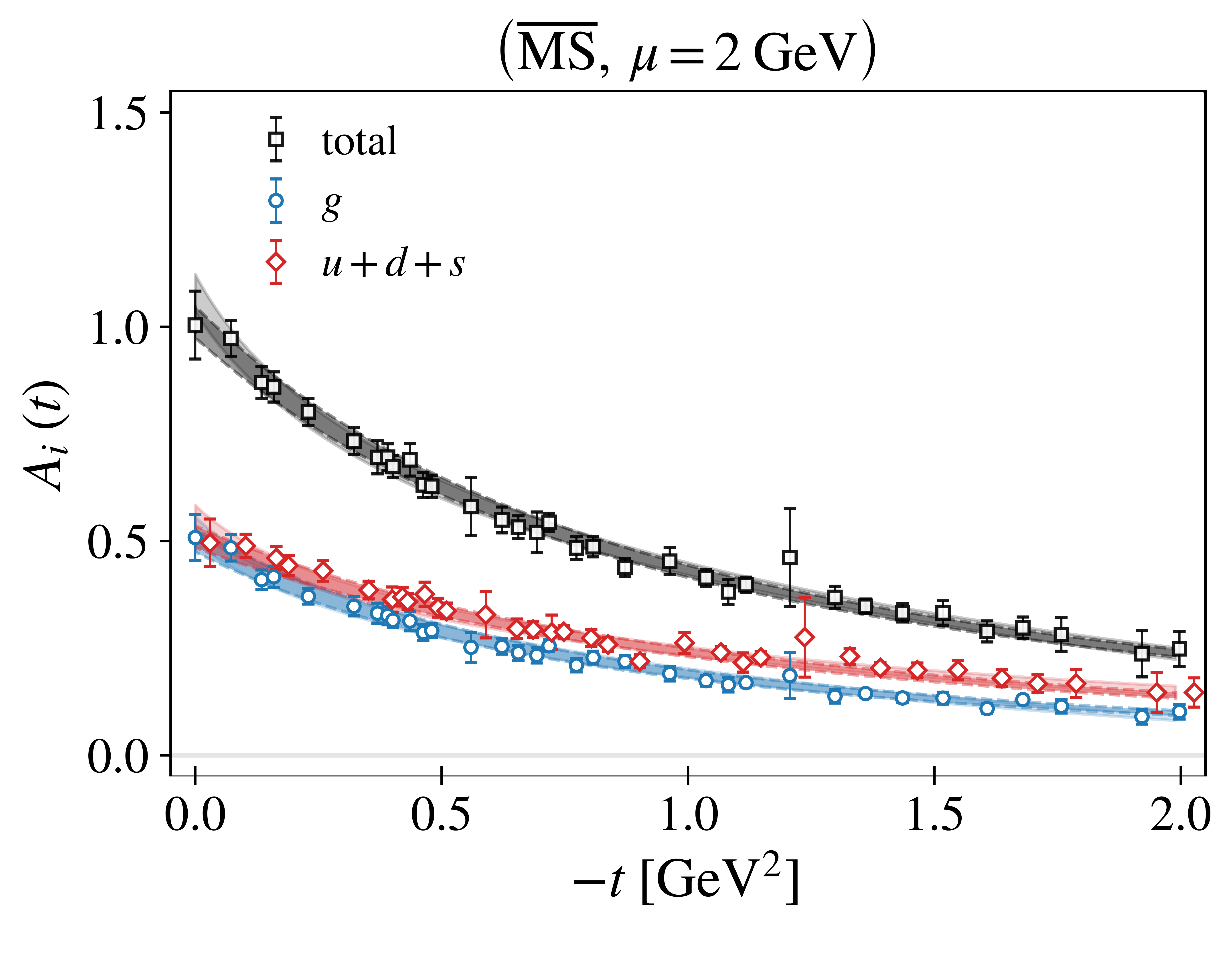}
    \caption{The gluonic and quarkonic GFF $A(t)$, describing the momentum distribution inside the proton. At rest the form factor sum up to 1, reproducing the expected momentum sum rule. Adapted from \cite{Hackett:2023rif}.}
    \label{fig:gff}
\end{figure}

The MICROSCOPE (Micro-Satellite à traînée Compensée pour l'Observation du Principe d’Équivalence) mission \cite{MICROSCOPE:2022doy, PhysRevLett.119.231101} has provided the most stringent empirical constraints to date on violations of the WEP, testing the universality of free fall for test masses in Earth's gravitational field. The result bounds any differential acceleration between test bodies to a precision of  
\begin{eqaed}
 \label{eq:weprocks}
\dfrac{m_g-m_i}{m_i}\lesssim 10^{-15} \, ,
\end{eqaed}
where $m_g,\,m_i$ denote gravitational and inertial mass respectively. This places severe limits on WEP-violating physics, and thus constrains the low-energy phenomenology of theories coupled to gravity.
Beyond macroscopic physics, the WEP can be tested for individual protons. In \cite{PhysRevLett.119.033001}, the inertial mass of the proton was determined experimentally using high-precision Penning trap mass spectrometry. By calibrating the measured cyclotron frequency of a single proton to that of a $^{12}C$ nucleus in the same trap, the proton-to-carbon mass ratio was obtained with an extremely high precision, leading to a proton mass of
\begin{eqaed}
 \label{eq:protonmassinert}
m_p= 0.93827198496\pm 3\times 10^{-11} \,\text{GeV} \, .    
\end{eqaed}
On the other hand, the gravitational mass of a nucleon, averaged over the proton and neutron masses, was obtained indirectly in \cite{Andreas_2011}. The number of atomic nuclei was obtained to high precision on a $\sim 1 $ kg sample of $^{28}Si$, which was then used to determine Avogadro's number. One can turn the logic around to determine the nucleon mass using the calibrated test masses and volumes, along with the molar mass measured by X-ray interferometry. In fact, one can obtain the precise mass of the $^{28}Si$ isotope, but the mass of the nucleons can only be obtained up the unaccounted binding energy. One has
\begin{eqaed}
m_{{^{28}}{Si}}= 26.0993547\pm 6\times10^{-7}\, \text{GeV} \, ,
\end{eqaed}
so that the upper bound $m_p\geq m_{{^{28}}{Si}}/28$ \footnote{The combined neutron-proton mass difference and binding energy is negative.} entails
\begin{eqaed}
\label{eq:protonmassgrav}
m_{p}\geq 0.9321198\pm 2\times10^{-8}\, \text{GeV} \, .
\end{eqaed}
Combining \cref{eq:protonmassgrav,eq:protonmassinert}, we see that the WEP for the proton must hold up to
\begin{eqaed}
\label{eq:wepprotons}
\dfrac{m_p^{(g)}-m_p^{(i)}}{m_p^{(i)}}\leq 7\times 10^{-3} \, .
\end{eqaed}
In order to estimate the gravitational and inertial masses of the proton in the dark bubble scenario we can rely on results from Lattice QCD (LQCD). This approach is particularly effective for studying low-energy QCD phenomena, such as hadron formation and confinement, which are typically inaccessible through perturbative methods. Within this framework it is possible to compute gravitational form factors (GFFs), matrix elements of the QCD energy-momentum tensor, $T^{\mu\nu}$. Analogously to electromagnetic form factors, GFFs encode information about the spatial distribution of mass, pressure, and shear forces within hadrons \cite{Hackett:2023nkr,Hackett:2023rif}. The decomposition takes the form
\begin{eqaed} 
\label{eq:EMT1}
&\langle N(\vec{p}\,', s') | \hat{T}^{\mu\nu} | N(\vec{p}, s) \rangle = \frac{1}{m} \, \bar{u}(\vec{p}\,', s') \bigg[ 
A(t)\, P^{\mu} P^{\nu} \\ 
&+J(t)\, i\, \sigma^{\rho\{\mu} P^{\nu\}} \Delta_{\rho} 
+ D(t)\, \frac{\Delta^{\mu} \Delta^{\nu} - g^{\mu\nu} \Delta^2}{4} 
\bigg] u(\vec{p}, s) \, .
\end{eqaed}
Recent advancements in LQCD allow for the decomposition of the proton's GFFs into contributions from up, down, strange quarks, and gluons. In \cite{Hackett:2023rif}, this flavor decomposition was performed over the range of momentum transfer $ 0 \leq -t \leq 2~\text{GeV}^2 $. For our purposes, up to a renormalization-group mixing \cite{Hackett:2023nkr} which is small enough not to affect our conclusions, the energy-momentum tensor admits a decomposition into quark and gluon contributions according to
\begin{eqaed}
    \hat{T}^{\mu\nu} = \hat{T}_q^{\mu\nu} + \hat{T}_g^{\mu\nu} \, ,
\end{eqaed}
with
\begin{eqaed}
\hat{T}_g^{\mu\nu} = 2\, \mathrm{Tr} \left[ F^{\mu\alpha} F^{\nu}_{\;\;\alpha} - \frac{1}{4} g^{\mu\nu} F^{\alpha\beta} F_{\alpha\beta} \right],\\
 \\
\hat{T}_q^{\mu\nu}= \sum_{f=u,d,s} \left[ \frac{i}{2} \bar{\psi}_f \left( \gamma^{\mu} D^{\nu} + \gamma^{\nu} D^{\mu} \right) \psi_f \right].
\end{eqaed}
Each GFF can then be decomposed as a gluon and quark contribution. These components are denoted $A(t)$, $J(t)$ and $D(t)$, as in \cref{eq:EMT1}, and they represent the momentum fraction, total angular momentum, and internal properties (e.g. pressure and shear) of the proton's constituents respectively. The forward limit $t = 0$ is of particular significance, since it describes the properties of proton structure at rest. LQCD results confirm that quark and gluon contributions to the total momentum fraction satisfy the expected momentum sum rule (up to uncertainties), as shown in \cref{fig:gff}. These lattice computations of GFFs have been compared to experiment for the case of diffractive photoproduction of $J/\psi$ \cite{GlueX:2023pev, Mamo:2022eui}, which is sensitive to the GFFs of nuclei. The GFFs of our interest from LQCD are given by \cite{Hackett:2023rif}
\begin{eqaed}
A_q(0) = 0.510\pm 0.025\, , \quad A_g(0) = 0.501\pm 0.027 \, ,
\end{eqaed}
which indeed are compatible with the sum rule $A_q(0) + A_g(0) = 1$,
in accordance with energy-momentum conservation. Likewise, the total angular momentum contributions obey the spin sum rule $J_q(0) + J_g(0) = \frac{1}{2}$, consistently with the fact that proton are fermions.
This shows that quarks and gluons contribute approximately equally to the proton’s momentum and spin. These results are also consistent with earlier LQCD studies \cite{Shanahan:2018pib}. 

At rest, in standard effective field theory the gravitational mass of the proton is given by
\begin{eqaed}
    \langle N| \hat{T}^{00} | N\rangle =A_g(0)m_p + A_q(0)m_p=m_p \, .
\end{eqaed}
However, as we discussed, in the dark bubble scenario gluons contribute with the opposite sign to the energy-momentum tensor up to subleading $\alpha'$ corrections, leading to
\begin{eqaed}
    \langle N | \hat{T}^{00}_q-\hat{T}^{00}_g | N \rangle =\left(0.009\pm 0.037 \right)m_p \, .
\end{eqaed}
The deviation between inertial and gravational masses is then
\begin{eqaed}
    \dfrac{m_p^{(g)}-m_p^{(i)}}{m_p^{(i)}}\simeq 0.99 \, ,
\end{eqaed}
which leads to a \textit{significant violation of the WEP} even for the most lax bounds in \cref{eq:wepprotons} coming from direct measurements of the proton mass. Extrapolating this deviation would also violate the macroscopic measurements in \cref{eq:weprocks} of the equivalence principle by 13 orders of magnitude.

\section{Conclusions}
The dark bubble scenario \cite{Danielsson:2009ff, Danielsson:2022lsl}, represents an alternative to standard 4d effective field theory with several attractive features. It avoids many arguments and no-go theorems that apply to standard compactifications. In particular, swampland conjectures are statements about effective field theories coupled to gravity (in the same dimensions), and they generally do not apply to the dark bubble. To wit, this scenario can easily realize (quasi-)de Sitter cosmologies, in contrast with the conjecture of \cite{Obied:2018sgi, Andriot:2018mav}, while also naturally accommodating chiral fermions and unitary gauge groups.

However, regardless of possible issues due to gravity being delocalized \cite{Banerjee:2020wov, Banerjee:2020wix}, non-abelian gauge fields couple to the induced gravity via higher-dimensional fields in a way that produces a negative sign at low energies. This leads to large violations of the WEP for hadrons, while the QGP thermal history and electroweak physics remain largely unaffected. Reconciling these predictions with existing data is extremely challenging: achieving compatibility would require fine-tuning effects of different orders in $\alpha'$, while maintaining their separation for all other observables. Given the empirical evidence constraining such deviations and the theoretical issues we discussed, we are led to conclude that the current formulation of the dark bubble scenario is incompatible with the Standard Model. To avoid this conclusion, a different mechanism without $B$ and $C$ fields must couple worldvolume fields and matter to the induced braneworld gravity. Presumably this would involve specific boundary conditions that realize the effective 4d graviton mode \cite{Banerjee:2020wix, Banerjee:2023uto}.

\section*{Acknowledgements}
We thank L. Bartolini, C. Bonanno, U. Danielsson, G. Dibitetto, S. Giri, D. Panizo and V. Van Hemelryck for insightful discussions.
\bibliographystyle{apsrev4-1}
\bibliography{bibliography}

\end{document}